\documentclass[sigconf,authorversion,nonacm]{acmart}
\AtBeginDocument{%
  \providecommand\BibTeX{{%
    \normalfont B\kern-0.5em{\scshape i\kern-0.25em b}\kern-0.8em\TeX}}}

\usepackage{algorithm2e}
\usepackage{booktabs}
\usepackage{subcaption}
\definecolor{apricot}{rgb}{0.98, 0.81, 0.69}
\definecolor{deeppeach}{rgb}{1.0, 0.8, 0.64}
\definecolor{lightpink}{rgb}{1.0, 0.71, 0.76}
\definecolor{pastelpink}{rgb}{1.0, 0.82, 0.86}
\definecolor{blizzardblue}{rgb}{0.67, 0.9, 0.93}
\usepackage[most]{tcolorbox}
\begin{document}

\title{Personal Research Knowledge Graphs}

\author{Prantika Chakraborty}
\affiliation{
\institution{Indian Association for the Cultivation of Science}
\city{Kolkata}
\country{India}
}
\email{prantika.ch@gmail.com}
\author{Sudakshina Dutta}
\affiliation{
\institution{Indian Institute of Technology Goa}
\city{Ponda}
\country{India}
}
\email{sudakshina@iitgoa.ac.in}

\author{Debarshi Kumar Sanyal}
\affiliation{
\institution{Indian Association for the Cultivation of Science}
\city{Kolkata}
\country{India}
}
\email{debarshi.sanyal@iacs.res.in}

\renewcommand{\shortauthors}{Chakraborty, Dutta, Sanyal}

\begin{abstract}
Maintaining research-related information in an organized manner can be challenging for a researcher. In this paper, we envision personal research knowledge graphs (PRKGs) as a means to represent structured information about the research activities of a researcher. PRKGs can be used to power intelligent personal assistants, and personalize various applications. We explore what entities and relations could be potentially included in a PRKG, how to extract them from various sources, and how to share a PRKG within a research group.
\end{abstract}




\maketitle

\section{Introduction}
Research is a complex activity; it requires not only a thorough understanding of the problem under consideration, familiarity with the prior art, and innovation of new ideas to tackle the problem but often, also managing a research lab or collaboration. To help researchers in their daily chores and their long term quest, advanced tools like academic search engines and recommendation systems are available. These tools may be standalone applications or tied to scholarly digital libraries like ACM Digital Library\footnote{https://dl.acm.org/} or even online social networks like ResearchGate\footnote{https://www.researchgate.net/}.  Although many of these tools can provide personalized recommendations, their model of a specific researcher is heavily dependent on how the researcher interacts with the tool, and it is often not transparent to the end-user. 
It is natural to expect that the greater the access to the personal information of the researcher, the better the tools perform. For example, consider the following imagined scenario where \textit{Sunita}, a computer scientist specialized in natural language processing (NLP), chats with her smart personal virtual assistant, \textit{SciJeeves}. The researcher had used an uncommon visualization tool to analyze a temporal data series. Sometime later, she is again required to visualize a time series but she cannot recollect the name of the visualization tool. She asks \textit{SciJeeves}: `What visualization tool did I use earlier to analyze time series?' and it immediately responds with the correct  name. The visualization tool, though uncommon or only privately available to the researcher, is frequently useful to her. Consider another query from the researcher: `How many machines in my lab have NVIDIA RTX 2000 series GPU?' \textit{SciJeeves}, being aware of the GPU configurations of the machines in the researcher's lab, is likely to offer a satisfactory response. Appendix \ref{appendix:A} shows more examples of conversations between \textit{Sunita} and \textit{SciJeeves}.

The above discussion motivates the need for a personal information system that captures the information that are relevant to the researcher rather than to the world at large. Knowledge graphs (KGs) \cite{gutierrez2021knowledge} are currently one of the most effective ways of organizing information and associated knowledge in the form of \textit{entities} and their \textit{relations}. Although there is a divergence of opinion regarding the definition of a KG \cite{ehrlinger2016towards,hogan2021knowledge,ji2021survey}, we follow a fairly inclusive one from \cite{ehrlinger2016towards}: 
Suppose we denote the set of entities by $\mathcal{E}$, and the set of relations by $\mathcal{R}$. A fact is a triple $(h, r, t)$ where $h \in \mathcal{E}$, $t \in \mathcal{E}$ and $r \in \mathcal{R}$. Let $\mathcal{F}$ denote the set of facts. Then, a knowledge graph is defined as $\mathcal{G} = \{\mathcal{E}, \mathcal{R}, \mathcal{F}\}$.

Existing KGs capture globally important objects like those found in Wikipedia\footnote{\url{https://en.wikipedia.org/wiki/Main_Page}} or  domain-specific resources like scholarly papers available on \texttt{arXiv}\footnote{\url{https://arxiv.org/}} or freely available COVID-19 related papers \cite{domingo2021covid}. But a scholarly knowledge graph built from a vast trove of research papers might not be sufficient to capture the information needs of a researcher. In the context of our examples above, a visualization tool that is privately available to the researcher and not mentioned in any paper is unlikely to be found in a public KG. Similarly, the configuration of the researcher's machine is not expected to be in a public KG. Even if the researcher uses a specific model like `Dell OptiPlex 7780' which might occur in a public KG, it does not reveal that the researcher's lab uses it because that information is not important to the larger user base of the KG. If the researcher is a reputed scientist, even small details about her lab might interest the public; however, her privacy concerns might inhibit indexing them in a public KG. Nonetheless, if a KG contained these information, it would be useful to a smart virtual assistant that she uses. 

To plug the gap between public KGs and a user-centric information system, Balog and Kenter \cite{balog2019personal} advance the notion of a \textit{personal knowledge graph (PKG)}, defining it as `a resource of structured information about entities personally related to its user, their attributes and the relations between them'.  Every node of a PKG is connected directly or indirectly to a central node that represents the user, leading to a `spider-web' structure of the graph. Balog and Kenter \cite{balog2019personal} cite the example of a PKG that stores entities like `acoustic-guitar' owned by the user and `Mom's dentist' whom the user's mother  consults. In contradistinction to a general-purpose KG, a PKG stores only those attributes of an entity (instead of all attributes) whose importance to the user is high. Further, entities as well as relations might be very short-lived. 

In this vision paper, we define a PKG specifically for researchers. 
We call such a KG a \textit{personal research knowledge graph (PRKG)}. The information captured by a PRKG are constrained to relate to her research activities only, rather than to her larger personal sphere. Note that for a researcher, research is likely to be a professional activity; our use of the term \textit{personal research} only intends to distinguish a given researcher's activities from those of others. 
By encoding personal information like her research interests, the computational tools used by her, and the specification of her lab equipment, a PRKG enables smart assistants to access a researcher's personal data to better respond to her research-related needs, and can even inspire new modes of researcher-machine interaction. PRKGs can be leveraged to personalize different scholarly applications like academic search engines (in customizing search results, query set expansion, etc.), recommendation systems (for papers, experts, and venues), and conversational chat-bots. When the user's personal data shared with these applications are sourced from a PRKG, the availability of the personal data is explainable to and controllable by the user. In this paper though we primarily consider PRKGs for computer scientists, researchers in other domains can also construct their own PRKGs.

In the rest of the article, we develop the novel idea of PRKGs in detail. In Section \ref{sec:RelatedWork}, we mention related works available in the literature. In Section \ref{sec:PRKGContent}, we identify potential content for a PRKG. In Section \ref{sec:PopulatingPRKG}, we discuss methods to populate it, and in Section \ref{sec:ExtendPRKG}, we briefly discuss how to share PRKGs within a research group. 
Section \ref{sec:Conclusion} concludes the paper.

\section{Related Work}
\label{sec:RelatedWork}
Existing literature related to this paper cuts across \textit{knowledge graphs}, \textit{personal knowledge graphs}, and \textit{the extraction of entities and relations from scientific documents}. Each of these areas is vast; so we will be brief and restrict the section to the research that is most closely connected with the present work.
\paragraph{Knowledge graphs}
KGs are increasingly used to represent \textit{knowledge} extracted from different sources; the representation takes the form of a graph -- its nodes represent entities and each of its labeled directed edges identifies a relation between a pair of entities. A lucid tutorial on KGs appears in \cite{hogan2021knowledge} and a recent survey is presented in  \cite{ji2021survey}. Open KGs such as DBpedia \cite{auer2007dbpedia}, YAGO \cite{suchanek2007yago}, ORKG \cite{jaradeh2019open} are publicly available and of utility to a broad user base, while enterprise KGs like those of Google, LinkedIn and Bloomberg are visible only within a company and cater to the needs of the respective business. Many open KGs are published as Linked Open Data RDF graphs that can be accessed in a variety of ways including through SPARQL queries. Recently developed KGs that are focused specifically in the area of scientific research and aim to exploit the vast amount of data contained in research publications include AI-KG \cite{aikg} and AIDA \cite{aida}. Active research areas in KGs involve knowledge acquisition (i.e., entity and relation extraction from different sources for KG construction, and completing KGs to predict missing links), and building knowledge-aware applications like question-answering over KGs \cite{ji2021survey}. 

\paragraph{Personal knowledge graphs}
PKGs are proposed in \cite{balog2019personal} as a structured model for organizing personal information of a user. The entities and relations in a PKG may not be of importance to other users. Open questions on PKGs include how to populate and maintain a PKG, how to evaluate it and how to utilize the knowledge captured in it.
Follow-up work by other authors addressed some of these questions in specific settings. Since entities relevant to a PKG are often present in personal conversations, researchers have proposed methods to extract such entities from conversations \cite{tigunova2020extracting, tigunova2021pride, joko2021conversational}. Due to low availability of training data for the task, these works advocate machine learning models that can learn from frugal training sets. 
Vannur et al. \cite{vannur2020data} discuss the difficulties of identifying fine-grained personal data entities (of types $birth\_year$, $gender$, etc.) and relations between persons in unstructured text corpora for PKG population, and demonstrate the use of rule-based annotators and a graph neural network for predicting missing links to populate a PKG from a given dataset. As regards the applications of PKGs, researchers have constructed PKGs to store health information of patients \cite{rastogi2020personal,ammar2021using} and planned to employ PKGs for various conversational systems \cite{balog2020common,balog2021conversational}.
\paragraph{Extracting entities and relations from scientific documents}  
The PRKG we propose incorporates scientific entities and relations extracted from various sources. Entity and relation extraction from natural language text is an active area of research  \cite{pawar2017relation,li2020survey}. Many recent works have focused on the scholarly domain. For example, given the abstract of an article,  joint entity and relation extraction frameworks have successfully used LSTMs \cite{luan2018multi}, dynamic span graphs \cite{luan2019general}, and pre-trained transformers \cite{wadden2019entity,eberts2019span,santosh2021joint}. Entity and relation extraction from full text of an article has also been considered \cite{jain-etal-2020-scirex,huang2021document,hou-etal-2021-tdmsci,viswanathan2021citationie} where the focus has been to extract tasks, methods, metrics, and datasets, using deep neural networks. 
However, extraction of scholarly entities and relations from private sources like emails, conversations, or social media posts remains less explored.

\begin{figure*}[!htbp]
    \centering
   \includegraphics[width=\textwidth]{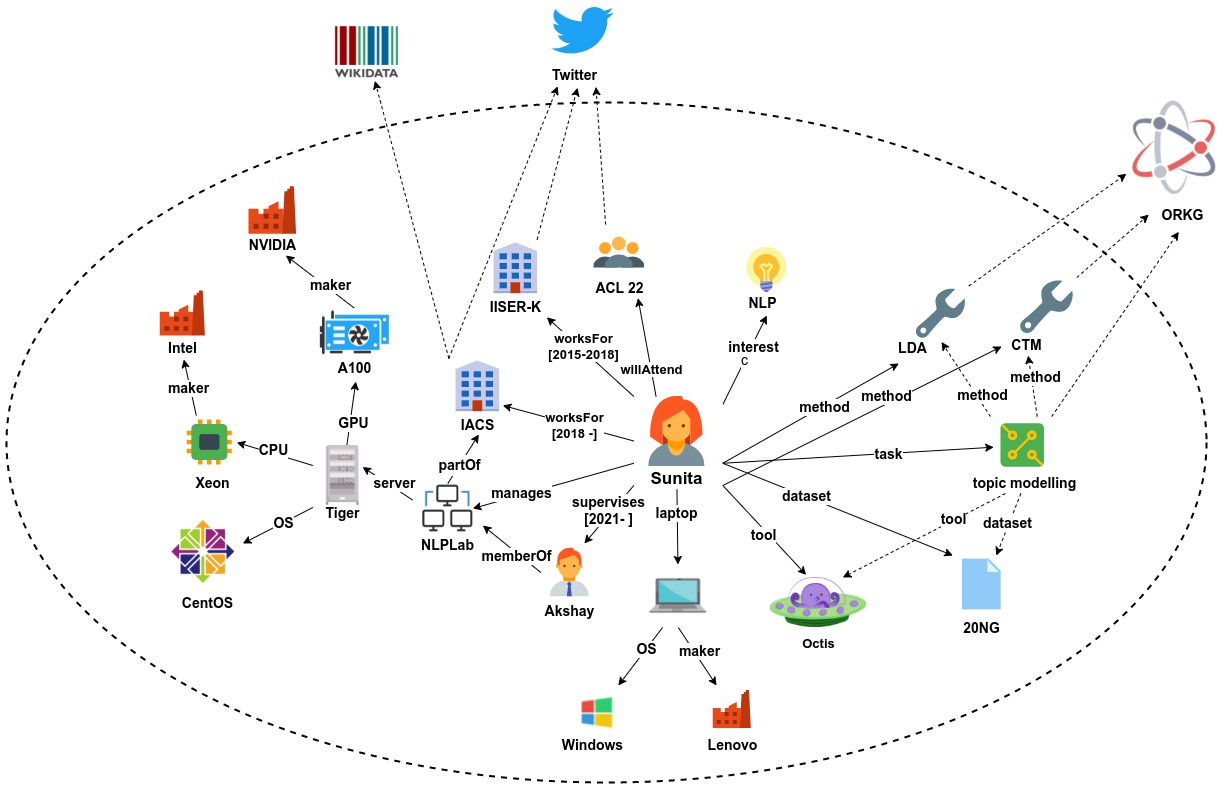}
    \Description[PRKG generated for a computer scientist \textit{Sunita}]{PRKG generated for a computer scientist \textit{Sunita} showing how she is connected to various research specific entities that are relevant to her}
    \caption{PRKG for a computer scientist \textit{Sunita}}
    \label{fig:prkg1}
\end{figure*}

\section{What's in a PRKG?}
\label{sec:PRKGContent}
\paragraph{Broad types of entities}
A PRKG belonging to a researcher contains information in the form of entities and their relationships that are of importance to the researcher. Entities in a PRKG may focus on the following aspects of the researcher: (1) Professional activities, (2) Personal as well as shared lab resources, (3) Fine-grained knowledge items related to her research.
Entities that identify her professional activities are her own current and past affiliations; her research interests;  her publications including papers, books and patents she authored; the talks she has given; the projects in which she participated / participates as principal investigator / co-principal investigator or in any other capacity; the conferences she plans to attend; the books or journals she is reading; the openings in her lab; and the courses she is currently offering. A researcher is likely to manage various equipment in her lab, and along with her personal resources, they may be included in the PRKG. 
Most importantly, the knowledge graph contains fine-grained knowledge entities relevant to the researcher and relations between them. For example, for an NLP researcher, it may include $(task, method, tool, dataset, metric)$ describing her current research task. Similarly, these related entities may be extracted from research papers authored or read by the researcher. Note that despite many recent initiatives like ORKG\footnote{\url{https://www.orkg.org/orkg/}}, OpenAlex\footnote{\url{https://openalex.org/}} and the now-retired Microsoft Academic Graph \cite{farber2019microsoft,herrmannova2016analysis}, we are not aware of any free open scholarly KG that contains fine-grained knowledge entities and relations (not just metadata) from a sufficiently large scholarly corpus that is very likely to include every paper being read by any researcher. However, there are many tools to parse scholarly papers. So, we may use them to extract entities and relations from the papers relevant to the researcher and add them to her PRKG. 

Most of the knowledge entities may be linked to external KGs like Wikidata\footnote{\url{https://www.wikidata.org/wiki/Wikidata:Main_Page}} or scholarly KGs like ORKG and AIKG\footnote{\url{http://w3id.org/aikg}}, and to online social networks. 
We understand that some of these relations are time-dependent. For example, since the researcher has switched jobs, the relation `worksFor' will have a start date and an end date. 
 One way to capture the temporal information is to augment each relation with temporal information. Symbolically, if $r \in \mathcal{R}$ is a relation, it may be augmented to `$r:[t_1, t_2]$' where $t_1$ and $t_2$ are start and end times (incorporated as attributes of the relation $r$), respectively, over which $r$ is valid. The end time $t_2$ is kept blank if the relation is still valid. 
 Entities no longer relevant may even be removed. A simple way to implement a PRKG is to model it as a labeled property graph in Neo4j\footnote{\url{https://neo4j.com/}}, and serialize it as an RDF graph for various downstream applications. Note that Neo4j is very flexible and allows one to create data (that is, nodes, relationships and properties) without defining a schema up front. 

\textit{An example:} 
Consider the PRKG shown in Fig. \ref{fig:prkg1} for a researcher \textit{Sunita}. She is an Assistant Professor at IACS since 2018 and has worked for IISER Kolkata earlier.  Note that `worksFor' is a relation which is annotated with a duration. Sunita's research interests are in NLP. Her current work on \textit{topic modeling} where she uses algorithms like latent Dirichlet allocation (LDA) and contextualized topic model (CTM) may be captured as the tuple $(task, method, tool, dataset, metric)$, which easily translates to triples in the PRKG. In other words, the PRKG includes triples of the form (`Sunita', {`task'}, $task$), 
(`Sunita', `method', $method$), 
(`Sunita', `tool', $tool$), 
 (`Sunita', `dataset', $dataset$), and (`Sunita', `metric', $metric$) where a relation (e.g., `tool') may be understood to be prefixed with `uses', which we omit for brevity. Since the method, tool, dataset and metric are related to the task in question, additional edges may be added to connect these entities to the task entity as shown by dotted lines in Figure \ref{fig:prkg1}. 
 Entities in the PRKG may be linked to corresponding entities in public knowledge bases like Wikidata (e.g., IACS corresponds to \url{www.wikidata.org/wiki/Q3347871}), social networks like Twitter\footnote{\url{https://twitter.com}}, and scholarly KGs like ORKG (e.g., LDA corresponds to \url{https://www.orkg.org/orkg/resource/R111035}). External knowledge bases and reasoners can be used to connect more entities like `pyLDAvis' (tool for topic  visualization) to the task `topic modeling'. However, unless \textit{Sunita} has used them, they will not be connected to her in the PRKG. Appendix \ref{appendix:B} details the implementation of a PRKG. 

\section{Populating a PRKG}
\label{sec:PopulatingPRKG}
The entities and the relations in the PRKG may be either curated manually by the researcher or extracted automatically using NLP techniques from structured or unstructured sources through a software agent. While the former is less prone to errors, it requires more human intervention and is, therefore, less scalable than the latter.
Information regarding her professional work including her affiliation, research interests, and publications may be primarily extracted from her curriculum vitae. Configuration of computers may be collected by installing an application in the respective devices.
Extracting knowledge entities (like tasks and methods), and collecting the details of other equipment and ongoing professional activities (like planned talks) are trickier. 
We discuss below some mechanisms to extract these entities and relations.

\paragraph{Conversations between researcher and chat-bot}
An AI chat-bot can mine personal information about a researcher by analyzing her conversions with it. For example, if the researcher utters informative statements like `I work in topic modeling' or `I used 20NG  dataset for topic modeling' in a chat, the agent should be able to parse the assertions and extract entities from them. However, it relies heavily on the user making personal assertions during conversations with the agent. If such assertions are absent or infrequent, extraction will suffer from low recall. In such cases, the chat-bot may try to elicit personal information  by proactively asking her probing questions like `What problem are you working on currently?' or `What software tools do you use in this task?'

\paragraph{Activity tracking}
An interesting approach could be to track the activities of the researcher, such as the papers she downloads or the queries she enters in scholarly search engines, and extract entities from them. 
To reduce noise, the agent might ask the researcher for the validity of the triples about which its confidence is low, and curate the valid ones in the graph. This is similar to how applications like Google Scholar and ResearchGate suggest edits to a user's profile. Textual sources like emails between the researcher and her group can also be used as information sources. The AI agent may also be deployed in 
passive listening mode when the researcher and her team participate in technical meetings, and entities and relations may be extracted from the utterances. We have observed that research meetings sometimes happen in informal settings in a mix of English and vernacular languages. This implies that the AI agent should be able to parse multilingual conversations. 
However, before adopting any of the above mechanisms, the privacy concerns of the researcher should be thoroughly considered.

\paragraph{Extraction from papers}
Recent papers authored by the researcher are authoritative sources of entities of her interest. But using only published works might delay the extraction of entities that relate to her ongoing research. To mitigate this problem, the chat-bot may be allowed to read the early drafts 
of her manuscripts. 
Unlike the first mechanism above, leveraging manuscripts as data source will allow the  identification of more fine-grained entities like performance scores,
keyphrases, keystone citations, and future work, without burdening the researcher with too many probing questions to be answered manually.  

Finally, symbolic and neural algorithms may be used to generate more relations among the existing entities and derive new knowledge from a PRKG \cite{chen2020review}. 

\section{Sharing a PRKG}
\label{sec:ExtendPRKG}
In many cases, especially in experimental sciences, research is a group activity. So a natural question is, can a researcher share her PRKG with other members of her group, while preferring to keep some of the information in it hidden from them?
One way is to incorporate role-based access control into the KG, where each group member will be assigned a \textit{role} which determines how they can access (read/write/append/control\footnote{\url{https://www.w3.org/wiki/WebAccessControl}}) the graph and its individual components (relations, node types, nodes, properties, etc.). For example, a researcher may share most of her PRKG with a collaborator but may not want the latter to know the papers she is reviewing. Following role-based access control (RBAC) model \cite{ferraiolo2003role}, a role \textit{`collaborator'} may be created and assigned \textit{`read'} access on the whole PRKG except the node denoting the paper to be reviewed. 
Support for fine-grained access control is limited in off-the-shelf graph databases; e.g., commercial versions of Neo4j support RBAC on the graph schema but not at the node instance level \cite{valzelli2020fine}. 

\section{Conclusion}
\label{sec:Conclusion}
We have presented personal research knowledge graphs as a structured organization of machine-actionable knowledge about a researcher. We have discussed the potential content of such a knowledge graph, how to populate it, and how to share it to a research group. We believe a PRKG will be extremely useful in designing personalized forms of knowledge-aware applications like scholarly search engines, conversational AI agents, and recommendation systems, that in turn can greatly assist a researcher in her everyday chores. However, accurate and timely collection of personally relevant entities and relations from diverse sources is a non-trivial challenge. 
Since a PRKG stores personal data, security and privacy must be given utmost priority when implementing, deploying, and sharing it.  
In the near future, we plan to design a framework with which researchers can create PRKGs easily. We also aim to implement an AI chat-bot powered by the knowledge in a PRKG. 
We hope this paper will spark further discussions on PRKGs and inform research on personalized virtual assistants for researchers.


\bibliographystyle{ACM-Reference-Format}
\bibliography{ref}


\begin{thebibliography}{36}


\ifx \showCODEN    \undefined \def \showCODEN     #1{\unskip}     \fi
\ifx \showDOI      \undefined \def \showDOI       #1{#1}\fi
\ifx \showISBNx    \undefined \def \showISBNx     #1{\unskip}     \fi
\ifx \showISBNxiii \undefined \def \showISBNxiii  #1{\unskip}     \fi
\ifx \showISSN     \undefined \def \showISSN      #1{\unskip}     \fi
\ifx \showLCCN     \undefined \def \showLCCN      #1{\unskip}     \fi
\ifx \shownote     \undefined \def \shownote      #1{#1}          \fi
\ifx \showarticletitle \undefined \def \showarticletitle #1{#1}   \fi
\ifx \showURL      \undefined \def \showURL       {\relax}        \fi
\providecommand\bibfield[2]{#2}
\providecommand\bibinfo[2]{#2}
\providecommand\natexlab[1]{#1}
\providecommand\showeprint[2][]{arXiv:#2}

\bibitem[Ammar et~al\mbox{.}(2021)]%
        {ammar2021using}
\bibfield{author}{\bibinfo{person}{Nariman Ammar}, \bibinfo{person}{James~E
  Bailey}, \bibinfo{person}{Robert~L Davis}, {and} \bibinfo{person}{Arash
  Shaban-Nejad}.} \bibinfo{year}{2021}\natexlab{}.
\newblock \showarticletitle{Using a Personal Health Library--Enabled {mHealth}
  Recommender System for Self-Management of Diabetes Among Underserved
  Populations: Use Case for Knowledge Graphs and Linked Data}.
\newblock \bibinfo{journal}{\emph{Journal of Medical Internet Research (JMIR)
  Formative Research}} \bibinfo{volume}{5}, \bibinfo{number}{3}
  (\bibinfo{year}{2021}), \bibinfo{pages}{e24738}.
\newblock


\bibitem[Angioni et~al\mbox{.}(2021)]%
        {aida}
\bibfield{author}{\bibinfo{person}{Simone Angioni}, \bibinfo{person}{Angelo
  Salatino}, \bibinfo{person}{Francesco Osborne}, \bibinfo{person}{Diego
  Reforgiato~Recupero}, {and} \bibinfo{person}{Enrico Motta}.}
  \bibinfo{year}{2021}\natexlab{}.
\newblock \showarticletitle{{AIDA}: a Knowledge Graph about Research Dynamics
  in Academia and Industry}.
\newblock \bibinfo{journal}{\emph{Quantitative Science Studies}}
  (\bibinfo{date}{11} \bibinfo{year}{2021}), \bibinfo{pages}{1--43}.
\newblock


\bibitem[Auer et~al\mbox{.}(2007)]%
        {auer2007dbpedia}
\bibfield{author}{\bibinfo{person}{S{\"o}ren Auer}, \bibinfo{person}{Christian
  Bizer}, \bibinfo{person}{Georgi Kobilarov}, \bibinfo{person}{Jens Lehmann},
  \bibinfo{person}{Richard Cyganiak}, {and} \bibinfo{person}{Zachary Ives}.}
  \bibinfo{year}{2007}\natexlab{}.
\newblock \showarticletitle{{DBpedia}: A nucleus for a web of open data}.
\newblock In \bibinfo{booktitle}{\emph{The Semantic Web}}.
  \bibinfo{publisher}{Springer}, \bibinfo{pages}{722--735}.
\newblock


\bibitem[Balog(2021)]%
        {balog2021conversational}
\bibfield{author}{\bibinfo{person}{Krisztian Balog}.}
  \bibinfo{year}{2021}\natexlab{}.
\newblock \showarticletitle{Conversational {AI} from an Information Retrieval
  Perspective: Remaining Challenges and a Case for User Simulation}. In
  \bibinfo{booktitle}{\emph{Proceedings of the Second International Conference
  on Design of Experimental Search {\&} Information REtrieval Systems, Padova,
  Italy, September 15-18, 2021}} \emph{(\bibinfo{series}{{CEUR} Workshop
  Proceedings}, Vol.~\bibinfo{volume}{2950})},
  \bibfield{editor}{\bibinfo{person}{Omar Alonso}, \bibinfo{person}{Stefano
  Marchesin}, \bibinfo{person}{Marc Najork}, {and} \bibinfo{person}{Gianmaria
  Silvello}} (Eds.). \bibinfo{pages}{80--90}.
\newblock


\bibitem[Balog et~al\mbox{.}(2020)]%
        {balog2020common}
\bibfield{author}{\bibinfo{person}{Krisztian Balog}, \bibinfo{person}{Lucie
  Flekova}, \bibinfo{person}{Matthias Hagen}, \bibinfo{person}{Rosie Jones},
  \bibinfo{person}{Martin Potthast}, \bibinfo{person}{Filip Radlinski},
  \bibinfo{person}{Mark Sanderson}, \bibinfo{person}{Svitlana Vakulenko}, {and}
  \bibinfo{person}{Hamed Zamani}.} \bibinfo{year}{2020}\natexlab{}.
\newblock \showarticletitle{Common Conversational Community Prototype:
  Scholarly Conversational Assistant}.
\newblock \bibinfo{journal}{\emph{arXiv preprint arXiv:2001.06910}}
  (\bibinfo{year}{2020}).
\newblock


\bibitem[Balog and Kenter(2019)]%
        {balog2019personal}
\bibfield{author}{\bibinfo{person}{Krisztian Balog} {and} \bibinfo{person}{Tom
  Kenter}.} \bibinfo{year}{2019}\natexlab{}.
\newblock \showarticletitle{Personal knowledge graphs: A research agenda}. In
  \bibinfo{booktitle}{\emph{Proceedings of the 2019 ACM SIGIR International
  Conference on Theory of Information Retrieval}}. \bibinfo{pages}{217--220}.
\newblock


\bibitem[Chen et~al\mbox{.}(2020)]%
        {chen2020review}
\bibfield{author}{\bibinfo{person}{Xiaojun Chen}, \bibinfo{person}{Shengbin
  Jia}, {and} \bibinfo{person}{Yang Xiang}.} \bibinfo{year}{2020}\natexlab{}.
\newblock \showarticletitle{A review: Knowledge reasoning over knowledge
  graph}.
\newblock \bibinfo{journal}{\emph{Expert Systems with Applications}}
  \bibinfo{volume}{141} (\bibinfo{year}{2020}), \bibinfo{pages}{112948}.
\newblock


\bibitem[Dessì et~al\mbox{.}(2020)]%
        {aikg}
\bibfield{author}{\bibinfo{person}{Danilo Dessì}, \bibinfo{person}{Francesco
  Osborne}, \bibinfo{person}{Diego Reforgiato~Recupero},
  \bibinfo{person}{Davide Buscaldi}, \bibinfo{person}{Enrico Motta}, {and}
  \bibinfo{person}{Harald Sack}.} \bibinfo{year}{2020}\natexlab{}.
\newblock \showarticletitle{{AI-KG}: An Automatically Generated Knowledge Graph
  of Artificial Intelligence}.
\newblock In \bibinfo{booktitle}{\emph{The Semantic Web}}.
\newblock


\bibitem[Domingo-Fern{\'a}ndez et~al\mbox{.}(2021)]%
        {domingo2021covid}
\bibfield{author}{\bibinfo{person}{Daniel Domingo-Fern{\'a}ndez},
  \bibinfo{person}{Shounak Baksi}, \bibinfo{person}{Bruce Schultz},
  \bibinfo{person}{Yojana Gadiya}, \bibinfo{person}{Reagon Karki},
  \bibinfo{person}{Tamara Raschka}, \bibinfo{person}{Christian Ebeling},
  \bibinfo{person}{Martin Hofmann-Apitius}, {and} \bibinfo{person}{Alpha~Tom
  Kodamullil}.} \bibinfo{year}{2021}\natexlab{}.
\newblock \showarticletitle{COVID-19 Knowledge Graph: a computable,
  multi-modal, cause-and-effect knowledge model of COVID-19 pathophysiology}.
\newblock \bibinfo{journal}{\emph{Bioinformatics}} \bibinfo{volume}{37},
  \bibinfo{number}{9} (\bibinfo{year}{2021}), \bibinfo{pages}{1332--1334}.
\newblock


\bibitem[Eberts and Ulges(2020)]%
        {eberts2019span}
\bibfield{author}{\bibinfo{person}{Markus Eberts} {and} \bibinfo{person}{Adrian
  Ulges}.} \bibinfo{year}{2020}\natexlab{}.
\newblock \showarticletitle{Span-based Joint Entity and Relation Extraction
  with Transformer Pre-training}.
\newblock \bibinfo{journal}{\emph{Proceedings of the 24th European Conference
  on Artificial Intelligence}}.
\newblock


\bibitem[Ehrlinger and W{\"o}{\ss}(2016)]%
        {ehrlinger2016towards}
\bibfield{author}{\bibinfo{person}{Lisa Ehrlinger} {and}
  \bibinfo{person}{Wolfram W{\"o}{\ss}}.} \bibinfo{year}{2016}\natexlab{}.
\newblock \showarticletitle{Towards a Definition of Knowledge Graphs.}
\newblock \bibinfo{journal}{\emph{SEMANTiCS (Posters and Demos)}}
  \bibinfo{volume}{48}, \bibinfo{number}{1-4} (\bibinfo{year}{2016}),
  \bibinfo{pages}{2}.
\newblock


\bibitem[F{\"a}rber(2019)]%
        {farber2019microsoft}
\bibfield{author}{\bibinfo{person}{Michael F{\"a}rber}.}
  \bibinfo{year}{2019}\natexlab{}.
\newblock \showarticletitle{The {Microsoft} academic knowledge graph: a linked
  data source with 8 billion triples of scholarly data}. In
  \bibinfo{booktitle}{\emph{ISWC}}. Springer, \bibinfo{pages}{113--129}.
\newblock


\bibitem[Ferraiolo et~al\mbox{.}(2003)]%
        {ferraiolo2003role}
\bibfield{author}{\bibinfo{person}{David Ferraiolo}, \bibinfo{person}{D~Richard
  Kuhn}, {and} \bibinfo{person}{Ramaswamy Chandramouli}.}
  \bibinfo{year}{2003}\natexlab{}.
\newblock \bibinfo{booktitle}{\emph{Role-based access control}}.
\newblock \bibinfo{publisher}{Artech house}.
\newblock


\bibitem[Guti{\'e}rrez and Sequeda(2021)]%
        {gutierrez2021knowledge}
\bibfield{author}{\bibinfo{person}{Claudio Guti{\'e}rrez} {and}
  \bibinfo{person}{Juan~F Sequeda}.} \bibinfo{year}{2021}\natexlab{}.
\newblock \showarticletitle{Knowledge graphs}.
\newblock \bibinfo{journal}{\emph{Commun. ACM}} \bibinfo{volume}{64},
  \bibinfo{number}{3} (\bibinfo{year}{2021}), \bibinfo{pages}{96--104}.
\newblock


\bibitem[Herrmannova and Knoth(2016)]%
        {herrmannova2016analysis}
\bibfield{author}{\bibinfo{person}{Drahomira Herrmannova} {and}
  \bibinfo{person}{Petr Knoth}.} \bibinfo{year}{2016}\natexlab{}.
\newblock \showarticletitle{An analysis of the {Microsoft} academic graph}.
\newblock \bibinfo{journal}{\emph{D-lib Magazine}} \bibinfo{volume}{22},
  \bibinfo{number}{9/10} (\bibinfo{year}{2016}), \bibinfo{pages}{37}.
\newblock


\bibitem[Hogan et~al\mbox{.}(2021)]%
        {hogan2021knowledge}
\bibfield{author}{\bibinfo{person}{Aidan Hogan}, \bibinfo{person}{Eva
  Blomqvist}, \bibinfo{person}{Michael Cochez}, \bibinfo{person}{Claudia
  d'Amato}, \bibinfo{person}{Gerard~de Melo}, \bibinfo{person}{Claudio
  Gutierrez}, \bibinfo{person}{Sabrina Kirrane}, \bibinfo{person}{Jos{\'e}
  Emilio~Labra Gayo}, \bibinfo{person}{Roberto Navigli},
  \bibinfo{person}{Sebastian Neumaier}, {et~al\mbox{.}}}
  \bibinfo{year}{2021}\natexlab{}.
\newblock \showarticletitle{Knowledge graphs}.
\newblock \bibinfo{journal}{\emph{Synthesis Lectures on Data, Semantics, and
  Knowledge}} \bibinfo{volume}{12}, \bibinfo{number}{2} (\bibinfo{year}{2021}),
  \bibinfo{pages}{1--257}.
\newblock


\bibitem[Hou et~al\mbox{.}(2021)]%
        {hou-etal-2021-tdmsci}
\bibfield{author}{\bibinfo{person}{Yufang Hou}, \bibinfo{person}{Charles
  Jochim}, \bibinfo{person}{Martin Gleize}, \bibinfo{person}{Francesca Bonin},
  {and} \bibinfo{person}{Debasis Ganguly}.} \bibinfo{year}{2021}\natexlab{}.
\newblock \showarticletitle{{TDMSci}: A Specialized Corpus for Scientific
  Literature Entity Tagging of Tasks Datasets and Metrics}. In
  \bibinfo{booktitle}{\emph{Proceedings of the 16th Conference of the European
  Chapter of the Association for Computational Linguistics: Main Volume}}.
  \bibinfo{publisher}{ACL}, \bibinfo{address}{Online},
  \bibinfo{pages}{707--714}.
\newblock


\bibitem[Huang et~al\mbox{.}(2021)]%
        {huang2021document}
\bibfield{author}{\bibinfo{person}{Kung-Hsiang Huang}, \bibinfo{person}{Sam
  Tang}, {and} \bibinfo{person}{Nanyun Peng}.} \bibinfo{year}{2021}\natexlab{}.
\newblock \showarticletitle{Document-level Entity-based Extraction as Template
  Generation}. In \bibinfo{booktitle}{\emph{Proceedings of the 2021 Conference
  on Empirical Methods in Natural Language Processing}}.
\newblock


\bibitem[Jain et~al\mbox{.}(2020)]%
        {jain-etal-2020-scirex}
\bibfield{author}{\bibinfo{person}{Sarthak Jain}, \bibinfo{person}{Madeleine
  van Zuylen}, \bibinfo{person}{Hannaneh Hajishirzi}, {and} \bibinfo{person}{Iz
  Beltagy}.} \bibinfo{year}{2020}\natexlab{}.
\newblock \showarticletitle{{S}ci{REX}: {A} Challenge Dataset for
  Document-Level Information Extraction}. In
  \bibinfo{booktitle}{\emph{Proceedings of the 58th Annual Meeting of the
  Association for Computational Linguistics}}. \bibinfo{publisher}{ACL},
  \bibinfo{address}{Online}, \bibinfo{pages}{7506--7516}.
\newblock


\bibitem[Jaradeh et~al\mbox{.}(2019)]%
        {jaradeh2019open}
\bibfield{author}{\bibinfo{person}{Mohamad~Yaser Jaradeh},
  \bibinfo{person}{Allard Oelen}, \bibinfo{person}{Kheir~Eddine Farfar},
  \bibinfo{person}{Manuel Prinz}, \bibinfo{person}{Jennifer D'Souza},
  \bibinfo{person}{G{\'a}bor Kismih{\'o}k}, \bibinfo{person}{Markus Stocker},
  {and} \bibinfo{person}{S{\"o}ren Auer}.} \bibinfo{year}{2019}\natexlab{}.
\newblock \showarticletitle{Open research knowledge graph: next generation
  infrastructure for semantic scholarly knowledge}. In
  \bibinfo{booktitle}{\emph{Proceedings of the 10th International Conference on
  Knowledge Capture}}. \bibinfo{pages}{243--246}.
\newblock


\bibitem[Ji et~al\mbox{.}(2021)]%
        {ji2021survey}
\bibfield{author}{\bibinfo{person}{Shaoxiong Ji}, \bibinfo{person}{Shirui Pan},
  \bibinfo{person}{Erik Cambria}, \bibinfo{person}{Pekka Marttinen}, {and}
  \bibinfo{person}{S~Yu Philip}.} \bibinfo{year}{2021}\natexlab{}.
\newblock \showarticletitle{A survey on knowledge graphs: Representation,
  acquisition, and applications}.
\newblock \bibinfo{journal}{\emph{IEEE Transactions on Neural Networks and
  Learning Systems}} (\bibinfo{year}{2021}).
\newblock


\bibitem[Joko et~al\mbox{.}(2021)]%
        {joko2021conversational}
\bibfield{author}{\bibinfo{person}{Hideaki Joko}, \bibinfo{person}{Faegheh
  Hasibi}, \bibinfo{person}{Krisztian Balog}, {and} \bibinfo{person}{Arjen~P.
  de Vries}.} \bibinfo{year}{2021}\natexlab{}.
\newblock \bibinfo{booktitle}{\emph{Conversational Entity Linking: Problem
  Definition and Datasets}}.
\newblock \bibinfo{publisher}{ACM}, \bibinfo{pages}{2390–2397}.
\newblock
\showISBNx{9781450380379}


\bibitem[Lee et~al\mbox{.}(2001)]%
        {cclinc}
\bibfield{author}{\bibinfo{person}{Young-Suk Lee}, \bibinfo{person}{Wu~Sok Yi},
  \bibinfo{person}{Stephanie Seneff}, {and} \bibinfo{person}{Clifford~J.
  Weinstein}.} \bibinfo{year}{2001}\natexlab{}.
\newblock \showarticletitle{Interlingua-Based Broad-Coverage
  {K}orean-to-{E}nglish Translation in {CCLINC}}. In
  \bibinfo{booktitle}{\emph{Proceedings of the First International Conference
  on Human Language Technology Research}}.
\newblock


\bibitem[Li et~al\mbox{.}(2020)]%
        {li2020survey}
\bibfield{author}{\bibinfo{person}{Jing Li}, \bibinfo{person}{Aixin Sun},
  \bibinfo{person}{Jianglei Han}, {and} \bibinfo{person}{Chenliang Li}.}
  \bibinfo{year}{2020}\natexlab{}.
\newblock \showarticletitle{A survey on deep learning for named entity
  recognition}.
\newblock \bibinfo{journal}{\emph{IEEE Transactions on Knowledge and Data
  Engineering}} (\bibinfo{year}{2020}).
\newblock


\bibitem[Luan et~al\mbox{.}(2018)]%
        {luan2018multi}
\bibfield{author}{\bibinfo{person}{Yi Luan}, \bibinfo{person}{Luheng He},
  \bibinfo{person}{Mari Ostendorf}, {and} \bibinfo{person}{Hannaneh
  Hajishirzi}.} \bibinfo{year}{2018}\natexlab{}.
\newblock \showarticletitle{Multi-Task Identification of Entities, Relations,
  and Coreference for Scientific Knowledge Graph Construction}. In
  \bibinfo{booktitle}{\emph{Proceedings of the 2018 Conference on Empirical
  Methods in Natural Language Processing}}. \bibinfo{pages}{3219--3232}.
\newblock


\bibitem[Luan et~al\mbox{.}(2019)]%
        {luan2019general}
\bibfield{author}{\bibinfo{person}{Yi Luan}, \bibinfo{person}{Dave Wadden},
  \bibinfo{person}{Luheng He}, \bibinfo{person}{Amy Shah},
  \bibinfo{person}{Mari Ostendorf}, {and} \bibinfo{person}{Hannaneh
  Hajishirzi}.} \bibinfo{year}{2019}\natexlab{}.
\newblock \showarticletitle{A general framework for information extraction
  using dynamic span graphs}. In \bibinfo{booktitle}{\emph{Proceedings of the
  2019 Conference of the North American Chapter of the Association for
  Computational Linguistics: Human Language Technologies, Volume 1 (Long and
  Short Papers)}}. \bibinfo{pages}{3036--3046}.
\newblock


\bibitem[Pawar et~al\mbox{.}(2017)]%
        {pawar2017relation}
\bibfield{author}{\bibinfo{person}{Sachin Pawar}, \bibinfo{person}{Girish~K.
  Palshikar}, {and} \bibinfo{person}{Pushpak Bhattacharyya}.}
  \bibinfo{year}{2017}\natexlab{}.
\newblock \showarticletitle{Relation Extraction : {A} Survey}.
\newblock \bibinfo{journal}{\emph{CoRR}}  \bibinfo{volume}{abs/1712.05191}
  (\bibinfo{year}{2017}).
\newblock
\showeprint[arXiv]{1712.05191}


\bibitem[Rastogi and Zaki(2020)]%
        {rastogi2020personal}
\bibfield{author}{\bibinfo{person}{Nidhi Rastogi} {and}
  \bibinfo{person}{Mohammed~J Zaki}.} \bibinfo{year}{2020}\natexlab{}.
\newblock \showarticletitle{Personal Health Knowledge Graphs for Patients}.
\newblock \bibinfo{journal}{\emph{arXiv preprint arXiv:2004.00071}}
  (\bibinfo{year}{2020}).
\newblock


\bibitem[Santosh et~al\mbox{.}(2021)]%
        {santosh2021joint}
\bibfield{author}{\bibinfo{person}{TYSS Santosh}, \bibinfo{person}{Prantika
  Chakraborty}, \bibinfo{person}{Sudakshina Dutta},
  \bibinfo{person}{Debarshi~Kumar Sanyal}, {and} \bibinfo{person}{Partha~Pratim
  Das}.} \bibinfo{year}{2021}\natexlab{}.
\newblock \showarticletitle{Joint Entity and Relation Extraction from
  Scientific Documents: Role of Linguistic Information and Entity Types}. In
  \bibinfo{booktitle}{\emph{Proceedings of the 2nd Workshop on Extraction and
  Evaluation of Knowledge Entities (EEKE), colocated with JCDL 2022}}.
\newblock


\bibitem[Suchanek et~al\mbox{.}(2007)]%
        {suchanek2007yago}
\bibfield{author}{\bibinfo{person}{Fabian~M Suchanek}, \bibinfo{person}{Gjergji
  Kasneci}, {and} \bibinfo{person}{Gerhard Weikum}.}
  \bibinfo{year}{2007}\natexlab{}.
\newblock \showarticletitle{{YAGO}: a core of semantic knowledge}. In
  \bibinfo{booktitle}{\emph{Proceedings of the 16th International Conference on
  World Wide Web}}. \bibinfo{pages}{697--706}.
\newblock


\bibitem[Tigunova(2020)]%
        {tigunova2020extracting}
\bibfield{author}{\bibinfo{person}{Anna Tigunova}.}
  \bibinfo{year}{2020}\natexlab{}.
\newblock \showarticletitle{Extracting personal information from
  conversations}. In \bibinfo{booktitle}{\emph{Companion Proceedings of the Web
  Conference 2020}}. \bibinfo{pages}{284--288}.
\newblock


\bibitem[Tigunova et~al\mbox{.}(2021)]%
        {tigunova2021pride}
\bibfield{author}{\bibinfo{person}{Anna Tigunova}, \bibinfo{person}{Paramita
  Mirza}, \bibinfo{person}{Andrew Yates}, {and} \bibinfo{person}{Gerhard
  Weikum}.} \bibinfo{year}{2021}\natexlab{}.
\newblock \showarticletitle{PRIDE: Predicting Relationships in Conversations}.
  In \bibinfo{booktitle}{\emph{Proceedings of the 2021 Conference on Empirical
  Methods in Natural Language Processing}}. \bibinfo{pages}{4636--4650}.
\newblock


\bibitem[Valzelli et~al\mbox{.}(2020)]%
        {valzelli2020fine}
\bibfield{author}{\bibinfo{person}{Marco Valzelli}, \bibinfo{person}{Andrea
  Maurino}, {and} \bibinfo{person}{Matteo Palmonari}.}
  \bibinfo{year}{2020}\natexlab{}.
\newblock \showarticletitle{A Fine-grained Access Control Model for Knowledge
  Graphs.}. In \bibinfo{booktitle}{\emph{17th International Conference on
  Security and Cryptography}}. \bibinfo{pages}{595--601}.
\newblock


\bibitem[Vannur et~al\mbox{.}(2021)]%
        {vannur2020data}
\bibfield{author}{\bibinfo{person}{Lingraj~S. Vannur}, \bibinfo{person}{Balaji
  Ganesan}, \bibinfo{person}{Lokesh Nagalapatti}, \bibinfo{person}{Hima Patel},
  {and} \bibinfo{person}{M.~N. Tippeswamy}.} \bibinfo{year}{2021}\natexlab{}.
\newblock \showarticletitle{Data Augmentation for Fairness in Personal
  Knowledge Base Population}. In \bibinfo{booktitle}{\emph{Pacific-Asia
  Conference on Knowledge Discovery and Data Mining (PAKDD Workshops)}}.
  \bibinfo{pages}{143--152}.
\newblock


\bibitem[Viswanathan et~al\mbox{.}(2021)]%
        {viswanathan2021citationie}
\bibfield{author}{\bibinfo{person}{Vijay Viswanathan}, \bibinfo{person}{Graham
  Neubig}, {and} \bibinfo{person}{Pengfei Liu}.}
  \bibinfo{year}{2021}\natexlab{}.
\newblock \showarticletitle{{CitationIE}: Leveraging the Citation Graph for
  Scientific Information Extraction}. In \bibinfo{booktitle}{\emph{Proceedings
  of the 59th Annual Meeting of the Association for Computational Linguistics
  and the 11th International Joint Conference on Natural Language Processing}}.
\newblock


\bibitem[Wadden et~al\mbox{.}(2019)]%
        {wadden2019entity}
\bibfield{author}{\bibinfo{person}{David Wadden}, \bibinfo{person}{Ulme
  Wennberg}, \bibinfo{person}{Yi Luan}, {and} \bibinfo{person}{Hannaneh
  Hajishirzi}.} \bibinfo{year}{2019}\natexlab{}.
\newblock \showarticletitle{Entity, Relation, and Event Extraction with
  Contextualized Span Representations}. In
  \bibinfo{booktitle}{\emph{Proceedings of the 2019 Conference on Empirical
  Methods in Natural Language Processing and the 9th International Joint
  Conference on Natural Language Processing (EMNLP-IJCNLP)}}.
  \bibinfo{pages}{5788--5793}.
\newblock


\end{thebibliography}

\appendix
\section{Appendix}
\subsection{Conversations with AI Chat-bot}
\label{appendix:A}
\begin{figure}[htbp]
\centering
    \subcaptionbox{Conversation 1: Sunita had worked on topic modeling and time-series analysis in a previous project. Now, she plans to work on a project on analysis of a news corpus. She chats with SciJeeves about the new project.\label{fig:convo1}}[1\columnwidth]{\includegraphics[width=1\columnwidth]{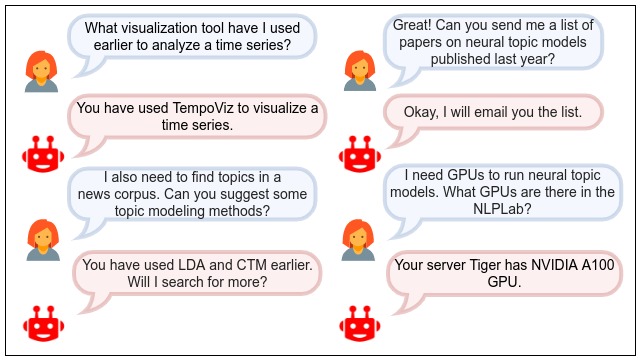}}
    \subcaptionbox{Conversation 2: Sunita chats with SciJeeves about the upcoming conferences in her field and her new project.\label{fig:convo2}}
    [1\columnwidth]{\includegraphics[width=1\columnwidth]{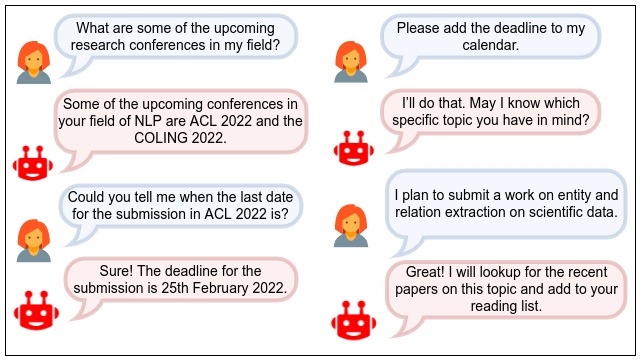}}
\Description[\textit{Sunita's} and \textit{SciJeeves'} conversations]{Conversations of a computer scientist, \textit{Sunita} and a personal virtual assistant for researchers, \textit{SciJeeves}}
\caption{Example conversations between \colorbox{cyan!10}{\textit{Sunita}}, a computer scientist, and \colorbox{pastelpink!50}{\textit{SciJeeves}}, a personalised virtual assistant for researchers.}
\label{fig:conversations}
\end{figure}

We present sample conversations between the computer scientist \textit{Sunita} and her conversational AI assistant \textit{SciJeeves} in Figures \ref{fig:convo1} and \ref{fig:convo2}. It shows how the knowledge in the PRKG maintained by Sunita is harnessed by SciJeeves to answer her queries and provide helpful suggestions. Note that these are \textit{imagined} dialogues mentioned to illustrate how an agent can help a researcher.

\subsection{Implementation of a PRKG}
\label{appendix:B}
\begin{figure}[!htbp]
   \includegraphics[width=0.96\linewidth]{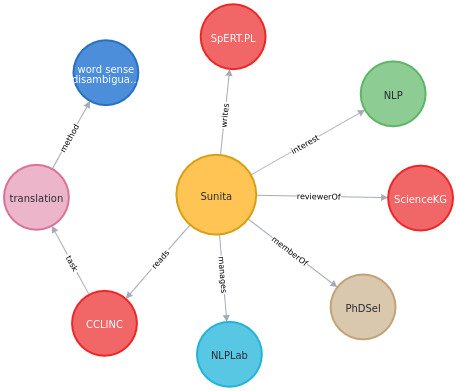}
   \Description[Neo4j implementation of compute scientist \textit{Sunita's} PRKG]{A Neo4j implementation of PRKG of a computer scientist \textit{Sunita} has been shown using entities and relations extracted from her personal resources.}
   \caption{PRKG for a computer scientist \textit{Sunita}, as implemented in Neo4j.}
    \label{fig:neo4j}
\end{figure}
We present the prototype of a PRKG in Fig. \ref{fig:neo4j} that we have implemented in the Enterprise edition of Neo4j Browser 4.4.0 included in Neo4j Desktop 1.4.10. The user of this PRKG is \textit{Sunita}, a computer scientist working in NLP. In order to populate this PRKG, we have taken a subset of the PRKG as described in Sec. \ref{sec:PRKGContent}, which includes the facts (`Sunita',  `interest',  `NLP') and (`Sunita',  `manages',  `NLPLab'). We have employed SpERT \cite{eberts2019span} for entity and relation extraction from scholarly papers. We trained SpERT on the SciERC \cite{luan2018multi} dataset and used it on the papers read by Sunita.  For example, we extracted the entities \textit{task} and \textit{method} from the statement  \textit{``High quality translation via word sense disambiguation and accurate word order generation of the target language.''} that appears in the abstract of the paper  \cite{cclinc}
They  are included in the PRKG as facts (`CCLINC',  `task',  `translation') and (`translation',  `method',  `word sense disambiguation'), where CCLINC\cite{cclinc} denotes the above paper. 
Fig. \ref{fig:neo4j} shows that Sunita is the writer of a paper labeled as `SpERT.PL' as shown by the triple (`Sunita', `writes', `SpERT.PL'), and reviewer of a paper `ScienceKG' as shown by (`Sunita', `reviewerOf', `ScienceKG'). She is also a member of a PhD selection committee as shown by (`Sunita', `memberOf', `PhDSel').

In order to make a provision for Sunita to share her PRKG with her collaborators and other research scholars in her group, we have used RBAC to restrict the collaborators' access to Sunita's confidential information. First, we have created a role \textit{`collaborator'} giving the collaborator all the access to the PRKG that an admin (here, Sunita) of the PRKG would have.

\[\texttt{CREATE ROLE collaborator AS COPY OF admin;}\]

Here \textit{admin} is an in-built role in Neo4j who is given complete access to the current graph database. But we have restricted the collaborator from viewing the nodes \textit{selComm} representing the selection committees (like the PhD Selection Committee) that she is a part of, using the following command in Neo4j:

    \[\texttt{DENY MATCH \{*\} ON GRAPH prkg  NODE selComm  TO collaborator;}\]

We have denied viewership of the papers that are being reviewed by Sunita, to the collaborator, using the following command: \begin{multline*}\texttt{DENY TRAVERSE ON GRAPH prkg RELATIONSHIP reviewerOf}\\ \texttt{TO collaborator;}
\end{multline*}

We have further made sure that the collaborator is unable to make any kind of modification, update or deletion to Sunita's PRKG:

\[\texttt{DENY WRITE \{*\} ON GRAPH prkg TO collaborator}\]

One restriction for the collaborator that we will like to bring into effect as part of the RBAC implemented here, is the prevention of \textit{`read'} access of the papers that are currently being written by Sunita but have not been accepted or published yet. This can be distinguished using the node level property \textit{status} that each \textit{paper} node will be assigned. This property can take the value of \textit{`published'}, \textit{`accepted'}, \textit{`underReview'} or \textit{`inProgress'}. Presently Neo4j does not allow the implementation of property level RBAC on nodes and relationships, and so we will be trying to implement it as a future work.

\end{document}